\documentclass[twocolumn,aps,showpacs,superscriptaddress]{revtex4}
\usepackage{graphics,graphicx,amsmath,amsfonts,amssymb,textcomp,wasysym}

\begin{document}


\title{A linear scaling DFT study of the growth of a new \{105\} facet layer on a Ge hut cluster}
\author{Sergiu Arapan}
\email{sergiu.arapan@gmail.com}
\affiliation{Computational Materials Science Unit, National Institute for Materials Science, 1-1 Namiki, Tsukuba, Ibaraki 305-0044, Japan}
\affiliation{Institute of Electronic Engineering and Nanotechnologies, Academy of Sciences of Moldova, Academiei 3/3, MD-2028 Chi\c{s}in\u{a}u, Moldova}
\author{David R. Bowler}
\email{david.bowler@ucl.ac.uk}
\affiliation{London Centre for Nanotechnology, University College London,17-19 Gordon Street, London WC1H 0AH, UK}
\affiliation{WPI-MANA, National Institute for Materials Science, Namiki 1-1, Tsukuba, 305-0044 Ibaraki, Japan}
\author{Tsuyoshi Miyazaki}
\email{MIYAZAKI.Tsuyoshi@nims.go.jp}
\affiliation{Computational Materials Science Unit, National Institute for Materials Science, 1-1 Namiki, Tsukuba, Ibaraki 305-0044, Japan}
\date{\today}


\begin{abstract}
Ge hut clusters, which appear during heteroepitaxy of Ge on Si(001), are prototypical examples of islands formed through strain.  Experimentally, complete facets are observed to form rapidly, though the mechanism is unknown.  We model the growth of new faces on Ge hut clusters, using linear scaling DFT.  We build realistic small huts on substrates, and show that the growth of \{105\} facets proceeds from top-to-bottom, even for these small huts.  The growth of the facet is driven by the reconstruction on the \{105\} facet, and nucleates at the boundaries of the facet.  We thus resolve any ambiguities in kinetic models of hut cluster growth.


\end{abstract}

\maketitle


Heteroepitaxy and strained growth have a long history, and are emerging as important techniques to improve the performance of microelectronic devices. In particular, the formation and stability of islands has been studied as a potential source of quantum dots, while nanowires are being pursued as potential channels for field-effect transistors~\cite{Aqua2013}. Germanium on silicon is a prototypical system, following a well-known progression from two dimensional growth, through the emergence of hut clusters and their transition into domes, but the structure and growth of these islands is poorly understood~\cite{Aqua2013,Mo1990,Voigtlander2001}.
Substantial efforts have been made to achieve a narrow distribution of these 3D islands by using kinetically self-limiting growth~\cite{Jesson1998,Kastner1999,McKay2008}. By assuming that islands grow by adding new facet layers, various kinetic models provide a fair description of experimental observations but cannot describe the actual mechanism of the facet growth. Direct STM imaging of 3D islands lacks any direct evidence for the facet nucleation sites and also cannot reveal the microscopic picture of the growth process.

However, an atomistic description of the Ge(105) surface may be provided by simulations. The first models for the Ge(105) surface reconstruction, as seen on hut cluster facets~\cite{Mo1990}, were given by density functional theory (DFT) calculations~\cite{Fujikawa2002,Raiteri2002}.  These showed that the basic unit of the reconstruction which is responsible for the stability of the surface was a U-shaped set of six atoms (known as Uss, and discussed further below).  Experimental and forcefield modelling~\cite{Montalenti2004} of the transition from the hut cluster to Ge domes showed that facets appear to nucleate at the apex, and that accumulation of partial facets near the top of large huts can drive the transition.  DFT modelling of the growth of the Ge(105) surface~\cite{Cereda2007} showed how new layers can nucleate, and drew on step-flow arguments to suggest reasons for the growth of facets from top to bottom.

Realistic modelling of the growth of facets, however, is challenging.  The reconstruction involves charge transfer between buckled Ge dimers~\cite{Fujikawa2002}, which requires a quantum mechanical method for accurate modelling.  Moreover, just as on the Si(001) surface, there are at least two ways to arrange the buckled dimers on the Ge(105) surface~\cite{Miyazaki2007}.  Conventional DFT does not allow simulations of realistic models of hut clusters (which will necessarily contain more than 10,000 atoms), while classical force-fields cannot account for quantum mechanical effects.  We have shown that linear scaling DFT methods~\cite{Bowler:2012zt} can be applied to model systems of over 2,000,000 atoms~\cite{Bowler:2010uq}.  We have applied these methods to models of small hut clusters (up to 23,000 atoms)~\cite{Miyazaki:2008tt}, showing that the transition from two-dimensional to three-dimensional growth is due to thermodynamic stability.

\begin{figure}[!h]
\begin{center}
\setlength\fboxsep{0pt}
\setlength\fboxrule{0.5pt}
\includegraphics[width=8cm]{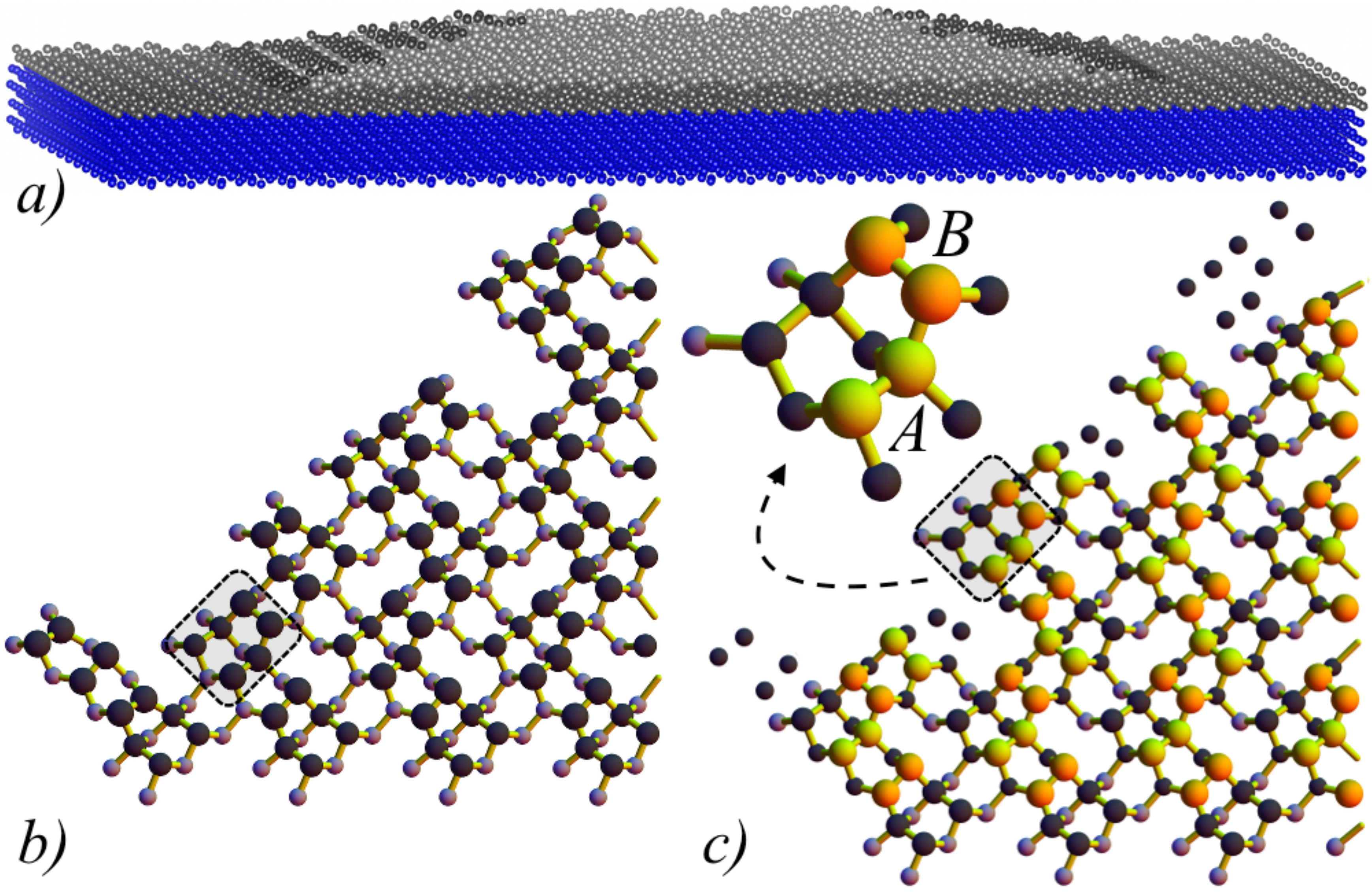}\\
\end{center}
\caption{(colour online) $a)$ A $16\times 26$ Ge hut (light- and dark-grey spheres) on a $24\times 36$ Si substrate (blue spheres) with two Ge wetting layers (grey spheres). Dark-grey spheres show atom positions of the outermost layer of small \{105\} facets. b) Top view of the left half of the left \{105\} facet of the hut shown above. Dark-grey spheres show positions of the \{105\} facet layer atoms (large spheres) as well as atoms of the (105) surface (smaller spheres) that form the Uss (marked with a grey box). c) Top view of the left half of the left \{105\} facet of the $16\times 28$ hut with additional facet layers (A-dimers and B-dimers are shown as yellow and orange spheres, respectively) and a close up view of a Uss. Atoms at positions close to that of atoms of the Uss of the previous facet (Fig.~\ref{fig1}b) are shown as dark-grey spheres.}
\label{fig1}
\end{figure}


In this work, we use linear scaling DFT methods to model the stability of dimers and groups of dimers, and the formation of a new layer, on the facet of a realistic Ge hut cluster.  We use the approach we have developed to model hut cluster structure\cite{Miyazaki:2008tt} to generate the starting hut cluster.  We notate the size of the substrate and the hut in terms of multiples of the cubic silicon (or germanium) bulk unit cell, as $L_{1}\times L_{2}$.  The starting system, shown in Fig.~\ref{fig1}a, consists of an elongated $16\times 26$ Ge hut on a $24\times 36$ Si substrate, with two layers of Ge as a wetting layer.  Both top and bottom of the substrate is reconstructed in the $p(2\times 2)$ reconstruction of Si(001) and Ge(001).  The system consists of 2,457 Ge atoms in the hut, 3,456 Ge atoms in the wetting layer and 13,824 Si atoms in the substrate, giving a total of 19,737 atoms.  To create the positions for dimers in the new facet, we also built a $16\times 28$ hut cluster, adding new facets on both ends of the hut (in this case with 19,973 atoms).  The size of the simulation cell was $24\times 36 \times 7.5$, giving a vacuum gap of $\sim 20$\AA.  The simulations used linear scaling DFT\cite{RevModPhys.71.1085,Bowler:2012zt} as implemented in the \textsc{Conquest} code\cite{Bowler:2010uq,Miyazaki:2004ee,Bowler:2002pt}, in the non-self-consistent Harris-Foulkes regime.  We used pseudo-atomic orbitals\cite{Torralba:2008wm} with single zeta basis functions in the local density approximation, which has been shown to model the Ge(105) surface and hut cluster facets well\cite{Miyazaki:2008tt,Miyazaki2007,Arita2014}.  The charge density grid used was equivalent to a 60\,Ha energy cutoff, and structural relaxation was performed via quenched molecular dynamics by using the FIRE method~\cite{Bitzek2006} with some modifications to account for slow convergence in the case of large systems~\cite{Arita2014}.

The structure of the \{105\} surface, particularly on the facet of a hut cluster, is based around arrangements of buckled Ge dimers.  We show the top two layers of atoms for part of a facet on our hut cluster model in Fig.~\ref{fig1}b.  The key building block of the surface is known as the Uss (U-shaped structure) and consists of three Ge dimers: an example is highlighted in a shaded rectangle in Fig.~\ref{fig1}b, and is enlarged in Fig.~\ref{fig1}c (where the three constituent dimers are the two labelled A and B and the dimer parallel to A).  We consider growth of the facet in terms of dimers, both because the Uss is built from dimers, and because the dimer is a very stable unit (ad-atoms diffuse quickly over the surface\cite{PhysRevB.70.245315}, and form dimers\cite{Cereda2007}).

The growth of a new facet will involve deposition of new dimers on the surface, continuing the \{105\} reconstruction.  Fig.~\ref{fig1}c shows the structure of a complete new facet, with the added dimers shown as yellow and orange spheres.  As can be seen in the shaded rectangle, two new dimers combine with dimers on the surface to create a new layer of Uss features, though it is important to note that this is altered at the edges of the facet (we return to this point later).  When considering the new dimers individually, however, it becomes clear that there are two types: A (coloured yellow); and B (coloured orange).  The A dimers are more stable than the B dimers on the perfect facet, as they can form relaxed bonds to the existing substrate atoms.  To form fully relaxed bonds, a B dimer must bond to an A dimer---that is, a dimer in the \emph{new} facet---as can be seen in the magnified view.  While B dimers can bond to the existing surface atoms, this leads to strained bonds and distorted bond angles.  The stable dimers formed from Ge adatoms during the growth of the perfect (105) surface\cite{Cereda2007} correspond to A dimers.


\begin{figure}[!h]
\begin{center}
\setlength\fboxsep{0pt}
\setlength\fboxrule{0.5pt}
\includegraphics[width=8cm]{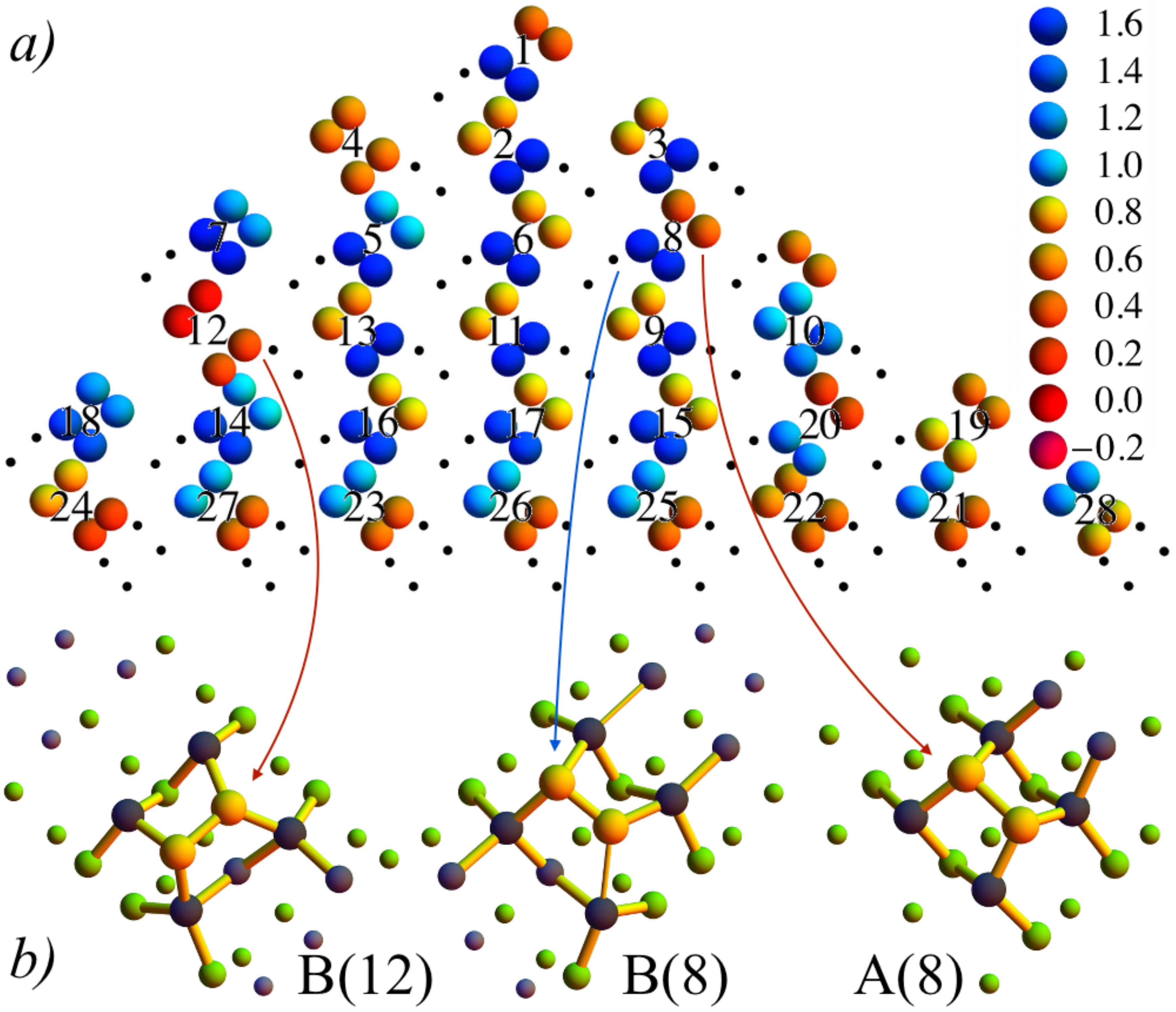}\\
\end{center}
\caption{(colour online) $a$) An energy map of single Ge ad-dimers: each pair of spheres shows the location of a single ad-dimer whose colour represents the ad-dimer energy. Small black spheres show the location of the third dimer that makes the Uss (this helps to identify B-dimers at the head of the Uss). Labels denote pairs of dimers on the same terrace. $b$) Local atomic structure (light green and dark grey spheres) of selected ad-dimers (yellow spheres). Dark grey spheres show optimised positions of atoms that form Uss of the underlying (105) surface. Bonds that exceed 2.8\AA~are shown as thin tubes.  }
\label{fig2}
\end{figure}

To understand the growth of a new facet on a realistic model of a hut cluster, we break down the problem: first, we will consider individual dimers on the facet; then we will examine interactions between pairs of dimers; finally, we will extend our studies to three dimers and then complete rows.  We show the energy for single dimers added to the small facet of the hut, relative to the energy of a dimer on the wetting layer, in Fig.~\ref{fig2}$a$ (studies of growth on the long facet will be presented in a future publication).  In the centre of the facet, it is clear that A-type dimers are more stable than B-type dimers, though the overall energy is slightly higher than on the wetting layer.  The structures of two typical dimers are shown in Fig.~\ref{fig2}$b$, where the stretched bond for the B-type dimer is shown as a thin bond.  In general, an A-type dimer inserts into the bonds of a single Uss on the surface, allowing it to make strong, relaxed bonds, while a B-type dimer bridges between two Uss, causing distorted, less stable bonds.

At the edges of the facet, however, the stability of dimers is very dependent on the local environment.  At the base of the facet, the first row of B-type dimers are very stable, due to the interface with the wetting layer.  There are several B-type dimers near the edge of the facet which are also stable, in particular B(12), which is illustrated in Fig.~\ref{fig2}$b$.  The stability of this dimer comes from the structure of the edge between facets, where there are broken bonds which can be passivated by the addition of the B-type dimer.  There are also A-type dimers which are significantly less stable than most, again due to changes of bonding at the edges between facets.

\begin{figure}[!h]
\begin{center}
\setlength\fboxsep{0pt}
\setlength\fboxrule{0.5pt}
\includegraphics[width=8cm]{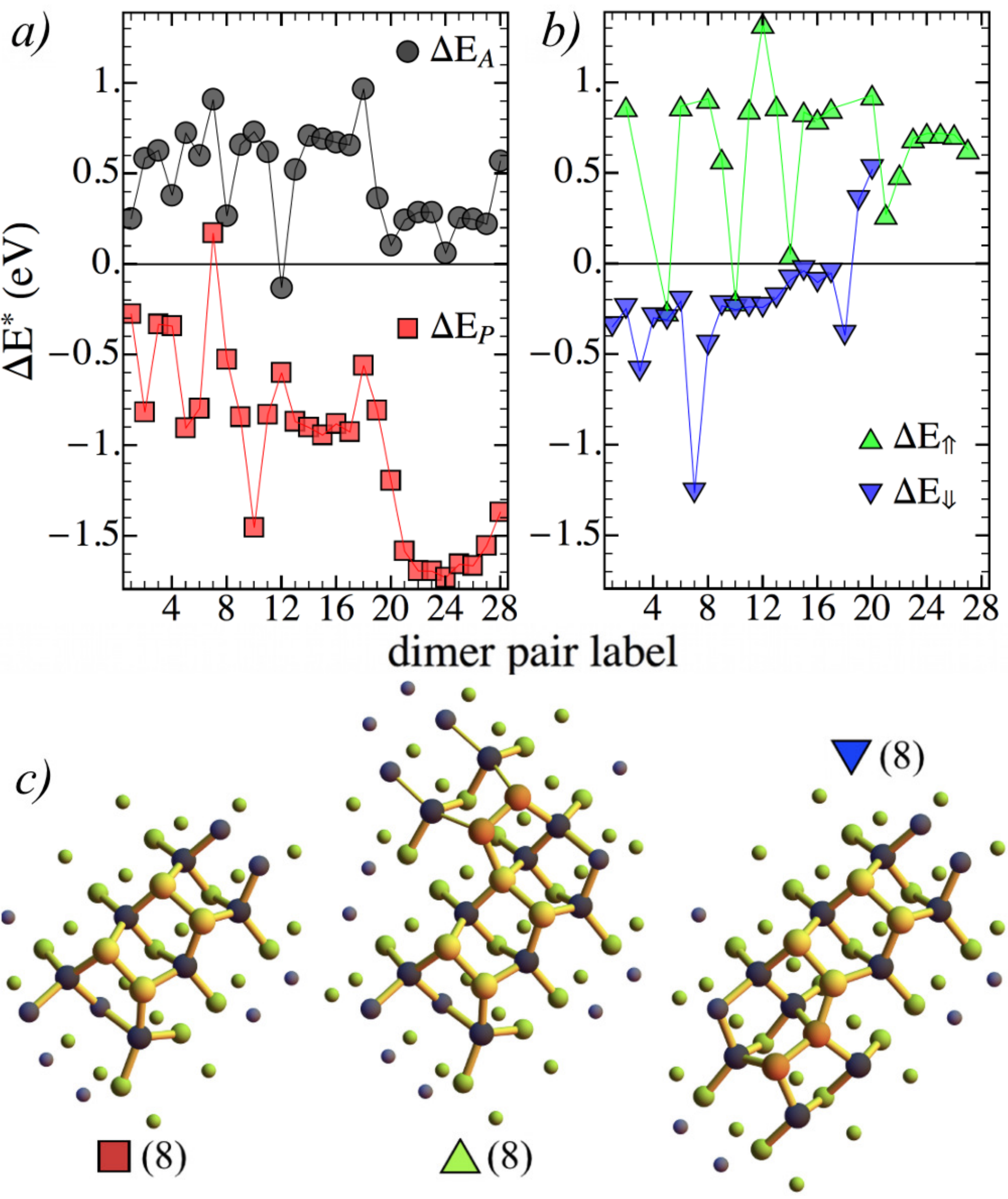}\\
\end{center}
\caption{(colour online) $a)$ Adsorption energy of single dimers added to either a clean facet (\CIRCLE), or an existing dimer on the facet to make a dimer pair ($\blacksquare$). The values relative to the adsorption energy of an ad-dimer on the wetting layer are shown. $b)$ Adsorption energy for a single dimer added to an existing dimer pair on the facet to make a 3-dimer structure ($\blacktriangle / \blacktriangledown$). $c)$ Local environment of a pair of ad-dimers (labeled 8, yellow spheres), and two 3-dimer structures, which have an ad-dimer (orange spheres) added upward/downward ($\blacktriangle / \blacktriangledown$) to that pair.}
\label{fig3}
\end{figure}

We plot the energy for individual dimers in Fig.~\ref{fig3}$a$.  There are several dimers within 0.2\,eV of the wetting layer energy, all A-type dimers associated with the edge or top of the facet (1, 8, 12, 20; we discount for now the dimers at the base, but will discuss them later).  These sites are candidates for the formation of the nucleus for a new facet.  Growth of the facet will proceed by addition of extra dimers to the stable nucleus, and we find that pairing one A-type and one B-type dimer gives a stable unit. An example of a stable pair is shown in Fig.~\ref{fig3}$c$ on the left, notated as $\blacksquare$(8), while the energy of all pairs in Fig.~\ref{fig2}$a$ is also plotted in Fig.~\ref{fig3}$a$, using the same numbering as before. Comparing the structures of dimers 8A and 8B in Fig.~\ref{fig2}$a$ with the dimer pair 8 in Fig.~\ref{fig3}$c$ shows the stabilisation mechanism: the B-type dimer passivates dangling bonds caused by the adsorption of the A-type dimer (and, equivalently, the A-type dimer reduces strain in the substrate atoms, releasing the strain in the B-type dimer bonds to the substrate).  
The energy of all pairs, except one, is lower than a corresponding pair on the wetting layer, giving a strong thermodynamic driving force for the growth of the hut.  While the total energy of some pairs is significantly lower than others, the location of the nucleation will depend on the \emph{first} dimer to adsorb.

Once a pair of dimers has formed, growth will proceed by adding a third dimer to the nucleus.  There are two ways to add another dimer, illustrated in Fig.~\ref{fig3}$c$: a B-type dimer can be added \emph{above} the dimer pair ($\blacktriangle$); or an A-type dimer can be added \emph{below} the dimer pair ($\blacktriangledown$).  Near the edges of the facet, however, there are restrictions: at the base, with the interface to the wetting layer, only B-type dimers can be added above; at the edges, with the interface to other facets, only A-type dimers can be added below. The change in energy for adding a dimer in both ways, at all sites on the facet, is plotted in Fig.~\ref{fig3}$b$.

It is striking that, for almost all sites, adding an A-type dimer below the pair is energetically favourable, while adding a B-type dimer above the pair is unfavourable, revealing a strong thermodynamic driving force for growth of a new facet from the top of the hut downwards.  In particular, adding new dimers above a pair at the base of the hut is energetically costly, while adding new dimers below existing pairs near the top of the facet is favourable, and more stable than adsorbing individual dimers on other parts of the facet.  The different behaviours of the dimers above and below the dimer pair is largely due to the difference in stability and bonding for A-type and B-type dimers.  However, the A-type dimer below the dimer pair also stabilises the dimer pair by completing the local structure of the Uss, while the B-type dimer above the dimer pair is bridging across a large gap caused by rebonding of the substrate atoms in the Uss above the dimer pair.  It is, in effect, forming part of two Uss features, both of which are incomplete and hence less stable than wetting layer dimers.
While the actual pathway for growth of a new facet of a hut cluster is likely to be highly complex, and to vary from facet to facet, we can say from these calculations that top-to-bottom growth is strongly favoured.  The adsorption sites for individual dimers, coupled with the difference in stability of dimers added above and below A-B dimer pairs, drive this energetic favourability.  

\begin{figure}[!h]
\begin{center}
\setlength\fboxsep{0pt}
\setlength\fboxrule{0.5pt}
\includegraphics[width=8cm]{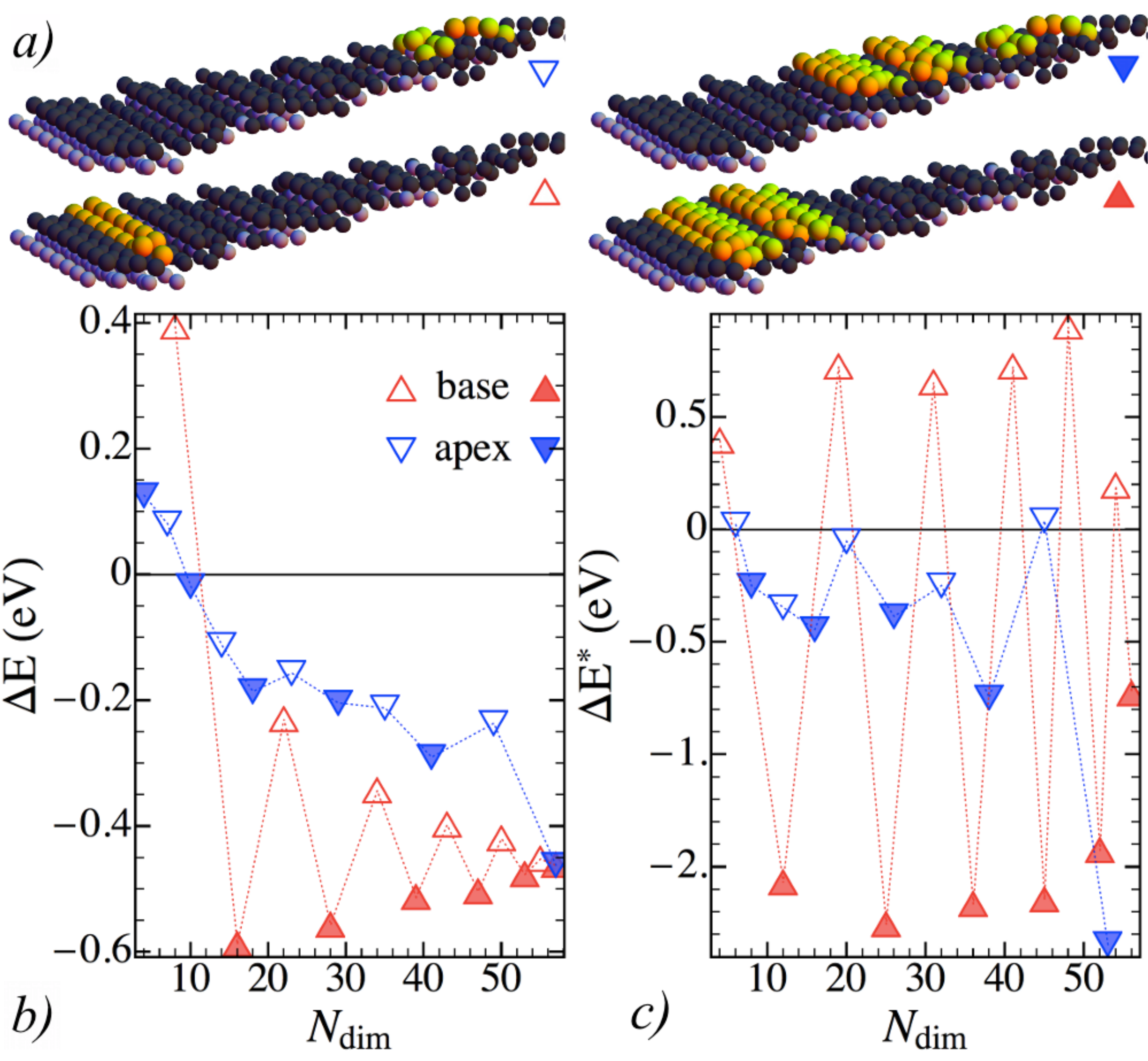}\\
\end{center}
\caption{(colour online) $a)$ Completion of \{105\} facet layers from base($\blacktriangle$)/apex($\blacktriangledown$) by adding rows of dimers one by one (consecutive rows are coloured in yellow and orange). $b)$ Energies (per ad-dimers) of various partial facet configurations ($\blacktriangle/\vartriangle$ - facet ending with a filled/half-filled terrace) with respect to the energy  of target hut. $c)$ Changes in energy/dimer for two consecutive partial facet configurations ($\blacktriangle/\vartriangle$ - filling/half-filling a terrace).}
\label{fig4}
\end{figure}

We test a somewhat simplified model of hut growth to explore further the energetics, completing the new facet row-by-row starting from either the base or the top of the hut, as illustrated in Fig.~\ref{fig4}$a$.  We study the energetics in two ways: first, by evaluating the energy per dimer of the partial facet (relative to the energy of a dimer on the wetting layer, as before), plotted in Fig.~\ref{fig4}$b$; second, by evaluating the \emph{change} in energy per dimer from row to row, plotted in Fig.~\ref{fig4}$c$.    Each row consists entirely of either A-type or B-type dimers, as can be seen in Fig.~\ref{fig2}$a$, and we consider two rows, one of A-type dimers and another of B-type dimers, to form a terrace.

Considering first the energy per dimer of the facet (Fig.~\ref{fig4}$b$), the growth from the top and the base show completely different behaviours.  The initial nucleation of the base is costly, while it is much smaller at the apex (particularly as this will involve fewer dimers).  Adding successive rows from the apex down gives a gradual and almost monotonic decrease in the energy, while the growth from the base oscillates, showing the high energetic cost for upward growth from dimer pairs seen in Fig.~\ref{fig3}.  This adds weight to the top-down growth model for the facet, as it will nucleate easily, and present a consistent, thermodynamically favourable energy surface for dimers on the facet.

This is further underlined when considering the change in energy from row to row (Fig.~\ref{fig4}$c$).  The growth from the base up will face a significant if not insuperable energetic barrier for each complete terrace (shown as open symbols).  By contrast, the growth from the apex downwards lowers the energy relative to the wetting layer with each row, whether starting or completing a terrace (with the possibly exception of two points where the energy change is essentially zero).  This clearly indicates that the downward growth of the facet is strongly favoured thermodynamically.  Of course, we have not provided a complete growth model, and it is not possible to do so.  However, this partial model clearly emphasises the key physics in the growth of new facets, and explains the experimentally observed trend for top-down growth.

In conclusion, we have used linear scaling DFT calculations of complete Ge hut clusters to explain why their facets grow from the top downwards.  Two key factors determine this growth direction.  First, the energetics of the underlying reconstruction, which favour the completion of single Uss structures on the surface over partial completion of pairs of Uss structures.  Second, the stability of dimers at the boundaries between facets and at the peak of the facet, which give stable nucleation sites for the new layer.  These insights require a number of capabilities only possible with linear scaling DFT: a realistic model of the complete hut; detailed structure of the facets and the edges between facets; and quantum mechanical modelling, which is crucial in determining the relative stabilities of different structures and reconstructions on semiconductor surfaces.  Given recent developments showing that highly scalable linear scaling molecular dynamics is possible\cite{Arita2014a}, we expect that applications of this technique will further insight into complex, important scientific problems.

We acknowledge Prof M. J. Gillan for helpful discussions and suggestions regarding calculations 
and interpretations of results.  
This work is partly supported by KAKENHI projects by MEXT (No. 22104005)
and JSPS (No. 26610120 and 26246021), Japan. 
The support from the Strategic Programs for Innovative Research (SPIRE) and the Computational Materials Science Initiative (CMSI) is also acknowledged.
Calculations were performed using 
K computer at RIKEN Advanced Institute for Computational Science(AICS), Kobe, Japan,
and the Numerical Materials Simulator at NIMS, Tsukuba, Japan.



\begin{thebibliography}{22}
\expandafter\ifx\csname natexlab\endcsname\relax\def\natexlab#1{#1}\fi
\expandafter\ifx\csname bibnamefont\endcsname\relax
  \def\bibnamefont#1{#1}\fi
\expandafter\ifx\csname bibfnamefont\endcsname\relax
  \def\bibfnamefont#1{#1}\fi
\expandafter\ifx\csname citenamefont\endcsname\relax
  \def\citenamefont#1{#1}\fi
\expandafter\ifx\csname url\endcsname\relax
  \def\url#1{\texttt{#1}}\fi
\expandafter\ifx\csname urlprefix\endcsname\relax\def\urlprefix{URL }\fi
\providecommand{\bibinfo}[2]{#2}
\providecommand{\eprint}[2][]{\url{#2}}

\bibitem[{\citenamefont{Aqua et~al.}(2013)\citenamefont{Aqua, Berbezier, Favre,
  Frisch, and Ronda}}]{Aqua2013}
\bibinfo{author}{\bibfnamefont{J.-N.} \bibnamefont{Aqua}},
  \bibinfo{author}{\bibfnamefont{I.}~\bibnamefont{Berbezier}},
  \bibinfo{author}{\bibfnamefont{L.}~\bibnamefont{Favre}},
  \bibinfo{author}{\bibfnamefont{T.}~\bibnamefont{Frisch}}, \bibnamefont{and}
  \bibinfo{author}{\bibfnamefont{A.}~\bibnamefont{Ronda}},
  \bibinfo{journal}{Phys. Rep.} \textbf{\bibinfo{volume}{522}},
  \bibinfo{pages}{59} (\bibinfo{year}{2013}).

\bibitem[{\citenamefont{Mo et~al.}(1990)\citenamefont{Mo, Savage,
  Swartzentruber, and Lagally}}]{Mo1990}
\bibinfo{author}{\bibfnamefont{Y.-W.} \bibnamefont{Mo}},
  \bibinfo{author}{\bibfnamefont{D.~E.} \bibnamefont{Savage}},
  \bibinfo{author}{\bibfnamefont{B.~S.} \bibnamefont{Swartzentruber}},
  \bibnamefont{and} \bibinfo{author}{\bibfnamefont{M.~G.}
  \bibnamefont{Lagally}}, \bibinfo{journal}{Phys. Rev. Lett.}
  \textbf{\bibinfo{volume}{65}}, \bibinfo{pages}{1020} (\bibinfo{year}{1990}).

\bibitem[{\citenamefont{Voigtl\"{a}nder}(2001)}]{Voigtlander2001}
\bibinfo{author}{\bibfnamefont{B.}~\bibnamefont{Voigtl\"{a}nder}},
  \bibinfo{journal}{Surf. Sci. Rep.} \textbf{\bibinfo{volume}{43}},
  \bibinfo{pages}{127} (\bibinfo{year}{2001}).

\bibitem[{\citenamefont{Jesson et~al.}(1998)\citenamefont{Jesson, Chen, Chen,
  and Pennycook}}]{Jesson1998}
\bibinfo{author}{\bibfnamefont{D.~E.} \bibnamefont{Jesson}},
  \bibinfo{author}{\bibfnamefont{G.}~\bibnamefont{Chen}},
  \bibinfo{author}{\bibfnamefont{K.~M.} \bibnamefont{Chen}}, \bibnamefont{and}
  \bibinfo{author}{\bibfnamefont{S.~J.} \bibnamefont{Pennycook}},
  \bibinfo{journal}{Phys. Rev. Lett.} \textbf{\bibinfo{volume}{80}},
  \bibinfo{pages}{5156} (\bibinfo{year}{1998}).

\bibitem[{\citenamefont{K\"{a}stner and Voigtl\"{a}nder}(1999)}]{Kastner1999}
\bibinfo{author}{\bibfnamefont{M.}~\bibnamefont{K\"{a}stner}} \bibnamefont{and}
  \bibinfo{author}{\bibfnamefont{B.}~\bibnamefont{Voigtl\"{a}nder}},
  \bibinfo{journal}{Phys. Rev. Lett.} \textbf{\bibinfo{volume}{82}},
  \bibinfo{pages}{2745} (\bibinfo{year}{1999}).

\bibitem[{\citenamefont{McKay et~al.}(2008)\citenamefont{McKay, Venables, and
  Drucker}}]{McKay2008}
\bibinfo{author}{\bibfnamefont{M.~R.} \bibnamefont{McKay}},
  \bibinfo{author}{\bibfnamefont{J.~A.} \bibnamefont{Venables}},
  \bibnamefont{and} \bibinfo{author}{\bibfnamefont{J.}~\bibnamefont{Drucker}},
  \bibinfo{journal}{Phys. Rev. Lett.} \textbf{\bibinfo{volume}{101}},
  \bibinfo{pages}{216104} (\bibinfo{year}{2008}).

\bibitem[{\citenamefont{Fujikawa et~al.}(2002)\citenamefont{Fujikawa, Akiyama,
  Nagao, Sakurai, Lagally, Hashimoto, Morikawa, and Terakura}}]{Fujikawa2002}
\bibinfo{author}{\bibfnamefont{Y.}~\bibnamefont{Fujikawa}},
  \bibinfo{author}{\bibfnamefont{K.}~\bibnamefont{Akiyama}},
  \bibinfo{author}{\bibfnamefont{T.}~\bibnamefont{Nagao}},
  \bibinfo{author}{\bibfnamefont{T.}~\bibnamefont{Sakurai}},
  \bibinfo{author}{\bibfnamefont{M.~G.} \bibnamefont{Lagally}},
  \bibinfo{author}{\bibfnamefont{T.}~\bibnamefont{Hashimoto}},
  \bibinfo{author}{\bibfnamefont{Y.}~\bibnamefont{Morikawa}}, \bibnamefont{and}
  \bibinfo{author}{\bibfnamefont{K.}~\bibnamefont{Terakura}},
  \bibinfo{journal}{Phys. Rev. Lett.} \textbf{\bibinfo{volume}{88}},
  \bibinfo{pages}{176101} (\bibinfo{year}{2002}).

\bibitem[{\citenamefont{Raiteri et~al.}(2002)\citenamefont{Raiteri, Migas,
  Miglio, Rastelli, and von K\"{a}nel}}]{Raiteri2002}
\bibinfo{author}{\bibfnamefont{P.}~\bibnamefont{Raiteri}},
  \bibinfo{author}{\bibfnamefont{D.~B.} \bibnamefont{Migas}},
  \bibinfo{author}{\bibfnamefont{L.}~\bibnamefont{Miglio}},
  \bibinfo{author}{\bibfnamefont{A.}~\bibnamefont{Rastelli}}, \bibnamefont{and}
  \bibinfo{author}{\bibfnamefont{H.}~\bibnamefont{von K\"{a}nel}},
  \bibinfo{journal}{Phys. Rev. Lett.} \textbf{\bibinfo{volume}{88}},
  \bibinfo{pages}{256103} (\bibinfo{year}{2002}).

\bibitem[{\citenamefont{Montalenti
  et~al.}(2004{\natexlab{a}})\citenamefont{Montalenti, Raiteri, Migas, von
  K\"{a}nel, Rastelli, Manzano, Costantini, Denker, Schmidt, Kern
  et~al.}}]{Montalenti2004}
\bibinfo{author}{\bibfnamefont{F.}~\bibnamefont{Montalenti}},
  \bibinfo{author}{\bibfnamefont{P.}~\bibnamefont{Raiteri}},
  \bibinfo{author}{\bibfnamefont{D.~B.} \bibnamefont{Migas}},
  \bibinfo{author}{\bibfnamefont{H.}~\bibnamefont{von K\"{a}nel}},
  \bibinfo{author}{\bibfnamefont{A.}~\bibnamefont{Rastelli}},
  \bibinfo{author}{\bibfnamefont{C.}~\bibnamefont{Manzano}},
  \bibinfo{author}{\bibfnamefont{G.}~\bibnamefont{Costantini}},
  \bibinfo{author}{\bibfnamefont{U.}~\bibnamefont{Denker}},
  \bibinfo{author}{\bibfnamefont{O.~G.} \bibnamefont{Schmidt}},
  \bibinfo{author}{\bibfnamefont{K.}~\bibnamefont{Kern}}, \bibnamefont{et~al.},
  \bibinfo{journal}{Phys. Rev. Lett.} \textbf{\bibinfo{volume}{93}},
  \bibinfo{pages}{216102} (\bibinfo{year}{2004}{\natexlab{a}}).

\bibitem[{\citenamefont{Cereda and Montalenti}(2007)}]{Cereda2007}
\bibinfo{author}{\bibfnamefont{S.}~\bibnamefont{Cereda}} \bibnamefont{and}
  \bibinfo{author}{\bibfnamefont{F.}~\bibnamefont{Montalenti}},
  \bibinfo{journal}{Phys. Rev. B} \textbf{\bibinfo{volume}{75}},
  \bibinfo{pages}{195321} (\bibinfo{year}{2007}).

\bibitem[{\citenamefont{Miyazaki et~al.}(2007)\citenamefont{Miyazaki, Bowler,
  Choudhury, and Gillan}}]{Miyazaki2007}
\bibinfo{author}{\bibfnamefont{T.}~\bibnamefont{Miyazaki}},
  \bibinfo{author}{\bibfnamefont{D.~R.} \bibnamefont{Bowler}},
  \bibinfo{author}{\bibfnamefont{R.}~\bibnamefont{Choudhury}},
  \bibnamefont{and} \bibinfo{author}{\bibfnamefont{M.~J.}
  \bibnamefont{Gillan}}, \bibinfo{journal}{Phys. Rev. B}
  \textbf{\bibinfo{volume}{76}}, \bibinfo{pages}{115327}
  (\bibinfo{year}{2007}).

\bibitem[{\citenamefont{Bowler and Miyazaki}(2012)}]{Bowler:2012zt}
\bibinfo{author}{\bibfnamefont{D.~R.} \bibnamefont{Bowler}} \bibnamefont{and}
  \bibinfo{author}{\bibfnamefont{T.}~\bibnamefont{Miyazaki}},
  \bibinfo{journal}{Rep. Prog. Phys.} \textbf{\bibinfo{volume}{75}},
  \bibinfo{pages}{36503} (\bibinfo{year}{2012}).

\bibitem[{\citenamefont{Bowler and Miyazaki}(2010)}]{Bowler:2010uq}
\bibinfo{author}{\bibfnamefont{D.~R.} \bibnamefont{Bowler}} \bibnamefont{and}
  \bibinfo{author}{\bibfnamefont{T.}~\bibnamefont{Miyazaki}},
  \bibinfo{journal}{J. Phys. Condens. Matter} \textbf{\bibinfo{volume}{22}},
  \bibinfo{pages}{74207} (\bibinfo{year}{2010}).

\bibitem[{\citenamefont{Miyazaki et~al.}(2008)\citenamefont{Miyazaki, Bowler,
  Gillan, and Ohno}}]{Miyazaki:2008tt}
\bibinfo{author}{\bibfnamefont{T.}~\bibnamefont{Miyazaki}},
  \bibinfo{author}{\bibfnamefont{D.~R.} \bibnamefont{Bowler}},
  \bibinfo{author}{\bibfnamefont{M.~J.} \bibnamefont{Gillan}},
  \bibnamefont{and} \bibinfo{author}{\bibfnamefont{T.}~\bibnamefont{Ohno}},
  \bibinfo{journal}{J. Phys. Soc. Jpn.} \textbf{\bibinfo{volume}{77}},
  \bibinfo{pages}{123706} (\bibinfo{year}{2008}).

\bibitem[{\citenamefont{Goedecker}(1999)}]{RevModPhys.71.1085}
\bibinfo{author}{\bibfnamefont{S.}~\bibnamefont{Goedecker}},
  \bibinfo{journal}{Rev. Mod. Phys.} \textbf{\bibinfo{volume}{71}},
  \bibinfo{pages}{1085} (\bibinfo{year}{1999}).

\bibitem[{\citenamefont{Miyazaki et~al.}(2004)\citenamefont{Miyazaki, Bowler,
  Choudhury, and Gillan}}]{Miyazaki:2004ee}
\bibinfo{author}{\bibfnamefont{T.}~\bibnamefont{Miyazaki}},
  \bibinfo{author}{\bibfnamefont{D.~R.} \bibnamefont{Bowler}},
  \bibinfo{author}{\bibfnamefont{R.}~\bibnamefont{Choudhury}},
  \bibnamefont{and} \bibinfo{author}{\bibfnamefont{M.~J.}
  \bibnamefont{Gillan}}, \bibinfo{journal}{J. Chem. Phys.}
  \textbf{\bibinfo{volume}{121}}, \bibinfo{pages}{6186} (\bibinfo{year}{2004}).

\bibitem[{\citenamefont{Bowler et~al.}(2002)\citenamefont{Bowler, Miyazaki, and
  Gillan}}]{Bowler:2002pt}
\bibinfo{author}{\bibfnamefont{D.~R.} \bibnamefont{Bowler}},
  \bibinfo{author}{\bibfnamefont{T.}~\bibnamefont{Miyazaki}}, \bibnamefont{and}
  \bibinfo{author}{\bibfnamefont{M.~J.} \bibnamefont{Gillan}},
  \bibinfo{journal}{J. Phys. Condens. Matter} \textbf{\bibinfo{volume}{14}},
  \bibinfo{pages}{2781} (\bibinfo{year}{2002}).

\bibitem[{\citenamefont{Torralba et~al.}(2008)\citenamefont{Torralba,
  Todorovic, Br\'{a}zdov\'{a}, Choudhury, Miyazaki, Gillan, and
  Bowler}}]{Torralba:2008wm}
\bibinfo{author}{\bibfnamefont{A.~S.} \bibnamefont{Torralba}},
  \bibinfo{author}{\bibfnamefont{M.}~\bibnamefont{Todorovic}},
  \bibinfo{author}{\bibfnamefont{V.}~\bibnamefont{Br\'{a}zdov\'{a}}},
  \bibinfo{author}{\bibfnamefont{R.}~\bibnamefont{Choudhury}},
  \bibinfo{author}{\bibfnamefont{T.}~\bibnamefont{Miyazaki}},
  \bibinfo{author}{\bibfnamefont{M.~J.} \bibnamefont{Gillan}},
  \bibnamefont{and} \bibinfo{author}{\bibfnamefont{D.~R.}
  \bibnamefont{Bowler}}, \bibinfo{journal}{J. Phys. Condens. Matter}
  \textbf{\bibinfo{volume}{20}}, \bibinfo{pages}{294206}
  (\bibinfo{year}{2008}).

\bibitem[{\citenamefont{Arita et~al.}(2014{\natexlab{a}})\citenamefont{Arita,
  Arapan, Bowler, and Miyazaki}}]{Arita2014}
\bibinfo{author}{\bibfnamefont{M.}~\bibnamefont{Arita}},
  \bibinfo{author}{\bibfnamefont{S.}~\bibnamefont{Arapan}},
  \bibinfo{author}{\bibfnamefont{D.~R.} \bibnamefont{Bowler}},
  \bibnamefont{and} \bibinfo{author}{\bibfnamefont{T.}~\bibnamefont{Miyazaki}},
  \bibinfo{journal}{J. Adv. Simul. Sci. Eng.} \textbf{\bibinfo{volume}{1}},
  \bibinfo{pages}{87} (\bibinfo{year}{2014}{\natexlab{a}}).

\bibitem[{\citenamefont{Bitzek et~al.}(2006)\citenamefont{Bitzek, Koskinen,
  G\"{a}hler, Moseler, and Gumbsch}}]{Bitzek2006}
\bibinfo{author}{\bibfnamefont{E.}~\bibnamefont{Bitzek}},
  \bibinfo{author}{\bibfnamefont{P.}~\bibnamefont{Koskinen}},
  \bibinfo{author}{\bibfnamefont{F.}~\bibnamefont{G\"{a}hler}},
  \bibinfo{author}{\bibfnamefont{M.}~\bibnamefont{Moseler}}, \bibnamefont{and}
  \bibinfo{author}{\bibfnamefont{P.}~\bibnamefont{Gumbsch}},
  \bibinfo{journal}{Phys. Rev. Lett.} \textbf{\bibinfo{volume}{97}},
  \bibinfo{pages}{170201} (\bibinfo{year}{2006}).

\bibitem[{\citenamefont{Montalenti
  et~al.}(2004{\natexlab{b}})\citenamefont{Montalenti, Migas, Gamba, and
  Miglio}}]{PhysRevB.70.245315}
\bibinfo{author}{\bibfnamefont{F.}~\bibnamefont{Montalenti}},
  \bibinfo{author}{\bibfnamefont{D.~B.} \bibnamefont{Migas}},
  \bibinfo{author}{\bibfnamefont{F.}~\bibnamefont{Gamba}}, \bibnamefont{and}
  \bibinfo{author}{\bibfnamefont{L.}~\bibnamefont{Miglio}},
  \bibinfo{journal}{Phys. Rev. B} \textbf{\bibinfo{volume}{70}},
  \bibinfo{pages}{245315} (\bibinfo{year}{2004}{\natexlab{b}}).

\bibitem[{\citenamefont{Arita et~al.}(2014{\natexlab{b}})\citenamefont{Arita,
  Bowler, and Miyazaki}}]{Arita2014a}
\bibinfo{author}{\bibfnamefont{M.}~\bibnamefont{Arita}},
  \bibinfo{author}{\bibfnamefont{D.~R.} \bibnamefont{Bowler}},
  \bibnamefont{and} \bibinfo{author}{\bibfnamefont{T.}~\bibnamefont{Miyazaki}},
  \bibinfo{journal}{J. Chem. Theory Comput.} \textbf{\bibinfo{volume}{10}},
  \bibinfo{pages}{5419} (\bibinfo{year}{2014}{\natexlab{b}}).

\end{thebibliography}

\end{document}